\documentclass[twocolumn,superscriptaddress,showpacs,nofootinbib,notitlepage,preprintnumbers,secnumarabic,amssymb, nobibnotes, aps, prd]{revtex4-2}
\usepackage[utf8]{inputenc}
\usepackage{graphicx}
\usepackage{latexsym,amsmath,amssymb,amsthm,lmodern,float,url,bm,IEEEtrantools}

\usepackage{listings}

\usepackage{mathtools}
\usepackage{natbib}
\usepackage{color}
\usepackage{microtype}
\usepackage{import}
\usepackage{bbold}
\usepackage[dvipsnames]{xcolor}
\usepackage[plain]{fancyref}
\usepackage{varioref}
\usepackage{slashed}
\usepackage{multirow}
\usepackage{tikz}

\usepackage{listings}
\usepackage{physics}
\usepackage{comment}
\usetikzlibrary{backgrounds}
\usetikzlibrary{arrows,shapes}
\usetikzlibrary{tikzmark}
\usetikzlibrary{calc}
\usetikzlibrary{positioning}
\usetikzlibrary{quantikz2}
\usepackage[normalem]{ulem}

\usetikzlibrary{positioning}
\usepackage{mathtools, nccmath}
\usepackage{wrapfig}
\usepackage{comment}
\usepackage{bbm}
\usepackage{adjustbox}

\usepackage[pdfusetitle]{hyperref}
\hypersetup{pdflang={English},colorlinks=true,linkcolor=RoyalBlue,citecolor=RoyalBlue,urlcolor=RoyalBlue} 
\usepackage[capitalise]{cleveref}

\usepackage{orcidlink}

\usepackage{todonotes}

\hypersetup{%
 bookmarksnumbered=true,
 pdftitle = {},
 pdfsubject = {},
 pdfauthor = {},
 pdfkeywords = {}
}

\newcommand{\fnal}{\affiliation{Fermi National Accelerator Laboratory, 
Batavia, IL 60510, USA}}

\newcommand{\epoq}{E$\rho$OQ}
\begin{document}

\preprint{FERMILAB-PUB-26-0155-T}
\title{Preparing Fermions 
via Classical Sampling and Linear Combinations of Unitaries}
\author{Erik J. Gustafson}
\email{egustafson@usra.edu}
\affiliation{Universities Space Research Association, Research Institute for Advanced Computer Science (RIACS), Mountain View, CA}
\author{Henry Lamm\,\orcidlink{0000-0003-3033-0791}}
\email{hlamm@fnal.gov}
\fnal
\begin{abstract}
    We present an extension of the Evolving density matrices 
on Qubits (E$\rho$OQ) framework that enables efficient fault-tolerant 
preparation of fermionic quantum states. The original method circumvents 
state preparation by stochastic sampling, but faces a sign problem in fermionic systems leading to a large number of circuits necessary. We resolve 
this by combining classical stochastic sampling with a linear combination 
of unitaries method that avoids the exponential circuit scaling that plagued na\"{i}ve implementations. The resulting algorithm requires $\mathcal{O}(M^2)$ $R_Z$ 
rotations for circuit preparation, where $M$ is the number of retained basis 
states. We validate the method for ground and excited states in the Thirring model, including by computing two-point correlation functions relevant to scattering.  In this model for fixed accuracy $\varepsilon$, $M$ is found to scale empirically as $M \propto \frac{1}{mg}\log(1/g)\log(1/m)$.
\end{abstract}
\maketitle

\textit{Introduction}.
State preparation remains a central open problem in quantum simulation, with rigorous complexity results establishing that it is QMA-hard~\cite{Kempe:2004sak}. In practice, end-to-end resource estimates for quantum field theory simulations frequently find that state preparation dominates e.g.~\cite{Jordan:2011ne,Lamm:2019uyc}, and this bottleneck persists across a range of relevant target states: vacuum, thermal, and scattering states. Identifying efficient and scalable methods is thus a prerequisite for realizing any quantum utility in subatomic physics and condensed matter.

There exist many proposals with varied performance and runtime guarantees. 
Adiabatic evolution \cite{Jordan:2011ne,Jordan:2014tma,Jordan:2017lea,Chakraborty:2020uhf,farhi2000quantum,Ingoldby:2025bdb} slowly evolves a known eigenstate of one Hamiltonian to another but requires long runtimes when crossing phase transitions or narrow energy gaps. 
Variational Quantum Eigensolvers (VQE)~\cite{Xie:2022jgj,Yamamoto:2021vxp,Popov:2023xft,Karalekas_2020,peruzzo2014variational,Davoudi:2025rdv,Chai:2025qhf,Davoudi:2024wyv}  and their extensions such as ADAPT \cite{Yao:2021ddt,Farrell:2023fgd,Ciavarella:2024lsp,Farrell:2024fit,Li:2025sgo,Grimsley:2018wnd,Gustafson:2024bww,Tang:2019tpm,Stadelmann:2025xjf,Zemlevskiy:2024vxt,Feniou:2023rtq}  and surrogate approaches \cite{Gustafson:2024bww,Gustafson:2024jop} can provide efficient circuits if the underlying NP-hard problem of optimizing their circuit is feasible and barren plateaus in the cost function landscape can be avoided. 

The Rodeo algorithm~\cite{Choi:2020tio} prepares energy eigenstates by repeatedly applying controlled time evolution to suppress contributions from unwanted states, with demonstrations on hardware~\cite{Qian:2021wya} and various model and modality extensions~\cite{Bazavov:2024spg,Rocha:2023bpe,Patkowski:2025leg,Rocha:2026ybx}. 
Quantum Krylov subspace methods~\cite{Candelori:2026jkr,Lee:2025mil,Kirby:2025iiw,Piccinelli:2025gmh,Oumarou:2025ypn} build a low-dimensional subspace from time-evolved states and diagonalize the Hamiltonian, avoiding difficulties of VQE while retaining shallow circuits. Wavelet-based methods~\cite{Brennen:2014iqu} encode momentum modes on separate registers, providing a natural renormalization-group structure that can reduce entanglement requirements. The Haag-Ruelle approach~\cite{Turco:2023rmx,Turco:2025jot} constructs asymptotic scattering states by smearing local operators against mode functions. The W state can be used as a resource for constructing wave packets with linear-depth circuits~\cite{Farrell:2025nkx}. Perturbative matching~\cite{Balaji:2025yua} uses weak-coupling expansions to construct an initial state close to the nonperturbative target, reducing the circuit depth needed by starting from a classical approximation. Other methods include simplified fermionic state preparation for NISQ devices~\cite{Hite:2025pvb}, dissipative cooling with provable convergence properties~\cite{Ding:2023ytq,Zhan:2025gof}. Another algorithm is Evolving density matrices On Qubits, or \epoq~\cite{Lamm:2018siq,Harmalkar:2020mpd,Gustafson:2020yfe,Saroni:2023uob}, which has seen some hardware demonstrations~\cite{Desaules:2023yhw}. This method, along with related work~\cite{Carena:2021ltu,Carena:2022hpz,Gupta:2025xti}, share the philosophy of leveraging classical Euclidean Monte Carlo to reduce the burden placed on the quantum device. \epoq~avoids the state-preparation problem entirely by classically sampling $\rho$ and passing superpositions of basis states onto quantum devices, and is particularly attractive for thermal~\cite{Harmalkar:2020mpd} or scattering states~\cite{Gustafson:2020yfe}. Gupta et al.~\cite{Gupta:2025xti} seeds the quantum evolution with an ansatz informed by Monte Carlo data.

The \epoq~algorithm is an $\mathcal{O}(1)$-time per circuit state preparation algorithm that prepares $M\ll 2^Q$ basis states on a $Q$ qubit quantum computer independently at the expense of needing $\mathcal{O}(M N_{shots}^2)$ circuits. It typically uses Quantum Monte Carlo methods to generate the initial $M$ configurations by sampling the density matrix weights $\rho_{mn}$. However the antisymmetry of fermionic wavefunctions introduces negative weights that cause a sign problem. Methods exist to ameliorate this fermionic sign problem such as reweighting schemes~\cite{Gibbs:1986ut}, fixed-node approximations~\cite{10.1063/1.443766,10.1063/5.0232424}, or contour deformations~\cite{Alexandru:2018ddf,Alexandru:2020wrj}. Alas, these either introduce uncontrolled systematics or retain a milder sign problem. Alternatively, by working in the $\beta \to \infty$ limit, stochastic sampling of $\rho$ can be replaced by a deterministic computation of the $M$ dominant amplitude contributions $\alpha_m$ via matrix product states (MPS) or density matrix renormalization group (DMRG)~\cite{stoudenmire2010minimally}. Quantum approaches suggest an alternative route: rather than modifying the sampling distribution, one can reformulate the sampling so that weights are obtained from quantum amplitudes~\cite{Tan:2022wqd} or eigenstate populations~\cite{Temme:2009wa}. These approaches effectively bypass the sign problem by retaining quantum coherence during sampling, rather than forcing a classical probabilistic representation, and may provide a path toward generating E$\rho$OQ input ensembles. In all approaches, another challenge remains: every basis state requires an independent quantum circuit so for fixed statistical precision we need $\mathcal{O}(M N_{shots}^2)$ total shots. In the presence of a sign problem with average sign $\langle\sigma\rangle$, the effective number of configurations grows as $\mathcal{O}(M\langle\sigma\rangle^{-1})$, giving a total cost of $\mathcal{O}(M \langle\sigma\rangle^{-1} N_{shots}^2)$. This potentially large number of circuits can be reduced by using the Linear Combination of Unitaries (LCU) method~\cite{Low:2016znh}.

LCU forms the backbone of our approach, which uses a combination of block-encoding techniques to ``load" a combination of Slater-determinants or configurations identified by classical methods as important states. This loading is done using the LCU method which has been successful in reducing computational costs in high energy physics \cite{Rhodes:2024zbr,Anderson:2024kfj,Kane:2024odt}, chemistry \cite{childs2012hamiltonian,Berry:2014ivo,Patel:2025nee,Georges:2024tzm}, linear systems solvers \cite{Childs:2015yvx,Chakraborty:2018cmg,Berry:2017fxp,Lee:2025vmt} and nonlinear systems \cite{liu2021efficient,childs2021high,Jennings:2025sqh}. 
This joint process changes the original gate depth scaling of \epoq~from $\mathcal{O}(1)$ to an $\mathcal{O}(M^2)$ process in terms circuit loading.

\textit{Theory.} Quantities of interest from quantum computers can be written as matrix elements of time-dependent operators $\mathcal{O}(t)$ between states $|\Psi_P\rangle $, $|\Psi_Q\rangle$. By applying projection operators $P$, $Q$ to a thermal state with nonzero overlap with them, it is possible to rewrite this matrix element in such a way to enable classical sampling from a density matrix $\rho$:

\begin{align}\label{eq:expectation}
\left<\Psi_P|\mathcal O(t)|\Psi_Q\right>
  &= \frac{\Tr P^{\dag} e^{-\beta H} Q \mathcal O(t)}{\Tr e^{-\beta H}}\equiv\frac{\langle Q\mathcal O(t) P^{\dag}\rangle_{\rho}}{\langle \delta_{mn}\rangle_{\rho}}
\end{align}

where $\langle\cdot\rangle_\rho$ denotes expectation values sampled from
$\rho_{mn}$, and the normalization $\langle\delta_{mn}\rangle_\rho$ can be
computed if needed~\cite{Harmalkar:2020mpd}. The weights $\rho_{mn}$ are
obtained via Euclidean path integral Monte Carlo with open boundary
conditions~\cite{Luscher:2011kk}, with configurations projected out by $P$, $Q$. Though $P$
and $Q$ are naturally handled classically, they can
alternatively be incorporated into the quantum circuit via projective
measurements~\cite{Barath}. Alas, $\mathcal{O}(t)$ introduces negativity into the classical sampling~\cite{Pashayan:2015cos} i.e. the real-time sign problem and therefore a classical computer alone is insufficient for these matrix elements. 

\epoq~offers a solution by using a hybrid approach: the classical computer obtains weights by
sampling, while a quantum computer determines matrix
elements between easily prepared basis states $|\psi_m\rangle$, avoiding the
need to directly prepare the target states $|\Psi_P\rangle $, $|\Psi_Q\rangle$
on the quantum device.
Classically, \epoq~draws samples from the density matrix
\begin{align}
    \rho =\ket{\psi}\bra{\psi}= e^{-\beta H} \propto \sum_{n,m} |\psi_n\rangle \langle \psi_m| \rho_{n,m}(\beta;H)
\end{align}
where $|n\rangle$ and $|m\rangle$ are bit strings and $\rho_{n,m}$ is the corresponding computational weight. 

On the quantum side, one must evaluate $\langle\psi_m|\mathcal{O}(t)|\psi_n\rangle$.
Since diagonal matrix elements $\langle\psi_m|\mathcal{O}(t)|\psi_m\rangle$ are
efficiently measurable on a quantum
processor~\cite{PhysRevLett.118.010501,PhysRevLett.123.070503,Roggero:2018hrn,Zohar:2018cwb,Clemente:2020lpr},
off-diagonal elements are recast using the superposition states
$|\psi_u\rangle, |\psi_v\rangle = |\psi_m\rangle \pm |\psi_n\rangle$, giving

\begin{equation}
    \langle \psi_m| \hat{\mathcal{O}} |\psi_n\rangle
    + \langle \psi_n| \hat{\mathcal{O}} |\psi_m\rangle
    = \langle \psi_u| \hat{\mathcal{O}} |\psi_u\rangle
    - \langle \psi_v| \hat{\mathcal{O}} |\psi_v\rangle.
\end{equation}
Thus, we need only prepare superpositions of basis state on the quantum computer. The full algorithm works by sampling bitstrings from a fast classical method and corresponding weights from a representation of an eigenstate, $|\bar{\Psi}\rangle$. A variety of other choices exist such as Markov chain Monte Carlo methods (MCMC), Selected Configuration Interaction (CI), DMRG \& projected entangled pair states (PEPS)~\cite{DeMeyer:2026ynw,Yang:2025rxv,Kelman:2024gjt,Emonts:2023ttz,Emonts:2018puo,Czarnik:2014gjj,Zohar:2017yxl,Kraus:2010zak}, or quantum-accelerated sampling. In this work, for simplicity we use DMRG~\cite{stoudenmire2010minimally}.
\begin{figure}[!t]
\includegraphics[width=\linewidth]{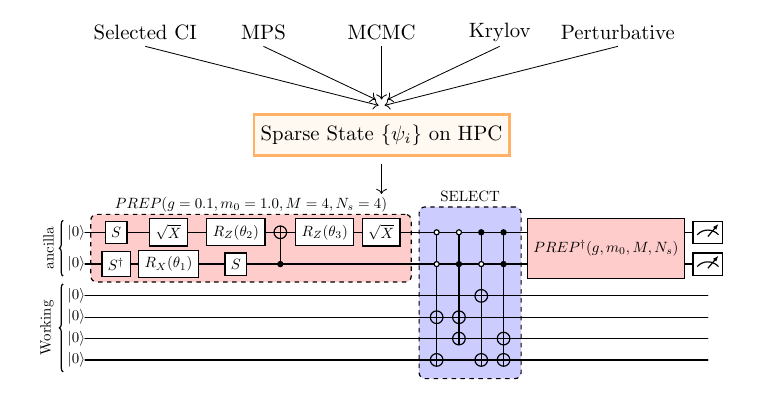}

\caption{Schematic implementation of the state preparation portion of proposed algorithm. It can take a classically-obtained fermionic initial state obtained from many sources, and compute time-evolved expectation values. The quantum circuit is for Eq.~\ref{eq:classic} with $\theta_1=0.57081,\theta_2=2.0663,\theta_3=0.62978$. The \textsc{Prep} circuit prepares the dense state while the \textsc{Select} circuit is comprised of multicontrolled gates which select the appropriate bit strings.}
\label{fig:algo_step}
\end{figure}

The modification of our algorithm works as follows:

\begin{enumerate}
    \item Using a classical method, generate a set of $M$ bitstrings $|\psi_m\rangle$ that form 
    the dominant contributions to an eigenstate $|\Psi\rangle$. For each bitstring, extract the corresponding probability 
    amplitude $\alpha_m = \langle \psi_m | \Psi\rangle$. 

    \item Construct a LCU circuit where \textsc{Prep} loads $\abs{\alpha_m}$ into an ancilla register, preparing the state
    \begin{equation}
        |\phi\rangle = \left(\sum_{m=1}^{M} |\alpha_m| |\psi_m\rangle\right)/ \left(\sum_{m=1}^{M} \alpha_m^2\right)^{-1/2}
    \end{equation}
    This requires $\lceil \log_2(M) \rceil$ ancilla qubits and 
    approximately $\mathcal{O}(M^2)$ $R_Z$ rotations. The success 
    probability of this preparation is $\|\alpha\|^2/\lambda^2$ where 
    $\lambda = \sum_i |\alpha_i|$.

    \item Use the ancilla register as a control for the \textsc{Select} 
    circuit, apply multicontrolled gates to the working 
    register initialized in $|0\rangle$ to load the corresponding 
    basis states. The phase information from the signs of 
    $\alpha_m$ is reintroduced at this stage.
\end{enumerate}

The benefit of this is that we develop a sampling improvement over standard  \epoq. By extracting approximate weights of the bit strings we can now load linear combination states and their normalization factors. Additional benefits are we can iteratively increase the bond-dimension of the $|\Psi\rangle$ to improve the accuracy of the state and perform extrapolations on the coefficients themselves as a function of the bond dimension to extend the reach beyond what might be feasible with a classical computer. 
The method for LCU preparation can be implemented through a variety of available packages such as those available in \textsc{Pennylane}~\cite{bergholm2018pennylane} or \textsc{Qiskit}~\cite{PhysRevA.93.032318,Javadi-Abhari:2024kbf}. 

As an example of the LCU method, consider the case of $m_0=1.0$, $g=0.1$, $N_s=4$ with $M=4$ computational states. The target initial state is
\begin{align}
\label{eq:classic}
    |\Psi\rangle &= \alpha |0101\rangle +\beta|0110\rangle+\beta |1001\rangle + \gamma |0011\rangle 
\end{align}
with $\alpha= -0.9346\hdots,\;
    \beta = -0.2117\hdots,\;
    \gamma= -0.1850\hdots$.
So on the ancilla register we want to prepare the state
\begin{align}
    |\phi\rangle = \alpha|00\rangle + \beta(|10\rangle + |01\rangle) + \gamma |11\rangle.
\end{align}
We show in Fig.~\ref{fig:algo_step} the LCU \textsc{Prep} and \textsc{Select} circuits for this example with a success probability $p_{suc}=77.2\%$. While this case may be high, $p_{suc}$ degrades as $M$ grows~\cite{Low:2016znh}: the 
measurement succeeds with probability $\|H|\psi\rangle\|^2/\lambda^2$, where 
$\lambda = \sum_i |\alpha_i|$ is the one-norm of the LCU coefficients. For $M$ 
bitstrings with roughly equal weights, $\lambda \sim \sqrt{M}$ and $p_{suc}=\mathcal{O}(1/M)$. As $M$ increases to faithfully represent the desired $\ket{\Psi}$, this  becomes a practical 
concern, and an analysis for the systems of interest will be important. One known remedy is oblivious amplitude amplification (OAA)~\cite{Berry:2013tiy,childs2012hamiltonian,Low:2016znh}, which can boost the 
success probability to $\mathcal{O}(1)$ at the cost of three additional queries to the \textsc{Prep} and \textsc{Select} oracles, 
provided the LCU coefficients are non-negative. This condition is satisfied 
since the $\alpha_i$ are quantum probability amplitudes.

 Keeping $M$ unique bitstrings requires $\lceil \log_2(M)\rceil$ qubits. This requires $2 (4^{\lceil \log_2(M)\rceil} - 1)\approx \mathcal{O}(M^2)$ $R_Z$ rotations for the \textsc{Prep} and \textsc{Select} circuits~\cite{Mottonen:2005mev}. In lattice field theories, we often encounter states with certain symmetries, e.g. baryon number or translational invariance. In these cases, there are fewer unique terms so we simulate a reduced number of basis states in the LCU scheme with increased amplitude~\cite{Saroni:2023uob}. For translational invariance, this reduces $M$ by a factor of $N^d$ with $N$ the number of sites and $d$ is the number of spatial dimensions. 
 
A special case of the \textsc{Prep} circuit arises when $\ket{\Psi}$ 
contains $M$ bitstrings of equal weight, as would generally occur via MCMC sampling of lattice field theories, where each configuration appears with equal 
frequency. In this case $\ket{\Psi}=\frac{1}{\sqrt{M}}\sum_m\ket{\psi_m}$ is a uniform superposition over a subset of the full $2^Q$ 
Hilbert space. Preparing such a state can be done using the 
quantum alias sampling construction of~\cite{Babbush:2018ywg} with 
$\mathcal{O}(M)$ gates and $\mathcal{O}(Q)$ ancilla qubits, or alternatively 
via the approach of~\cite{Shende:2006onn} at a cost of $\mathcal{O}(2^Q)$ gates, 
which becomes prohibitive for large $Q$.

\textit{Systematic error in truncation.} The state loaded onto the quantum computer, 
$|\phi\rangle$, approximates the target state as $|\Psi\rangle = \alpha |\phi\rangle + \varepsilon|\phi^\perp\rangle$, where $|\alpha|^2 + |\varepsilon|^2 = 1$ and $|\phi^\perp\rangle$ is the unimplemented 
component. The expectation value of any observable then decomposes as

\begin{align}
   \langle \Psi|O|\Psi\rangle = |\alpha|^2 \langle \phi|O|\phi\rangle 
   + |\varepsilon|^2 \langle \phi^\perp|O|\phi^\perp\rangle \notag\\ 
   + 2\,\mathrm{Re}\!\left(\alpha \varepsilon^* \langle \phi^\perp|O|\phi\rangle\right).
\end{align}

Applying the triangle inequality with $|\langle\phi|O|\chi\rangle| \leq \|O\|_2$ 
to the difference $|\langle \Psi|O|\Psi\rangle - |\alpha|^2\langle\phi|O|\phi\rangle|$, 
the systematic uncertainty is bounded by

\begin{align}
    \delta O_{\mathrm{sys}} \leq \left(|\varepsilon|^2 + 2|\varepsilon||\alpha|\right)\|O\|_2,
\end{align}

where $\|O\|_2$ is the operator 2-norm. This error is improvable by increasing 
the bond dimension or $M$.

\begin{figure}
    \includegraphics[width=\linewidth]{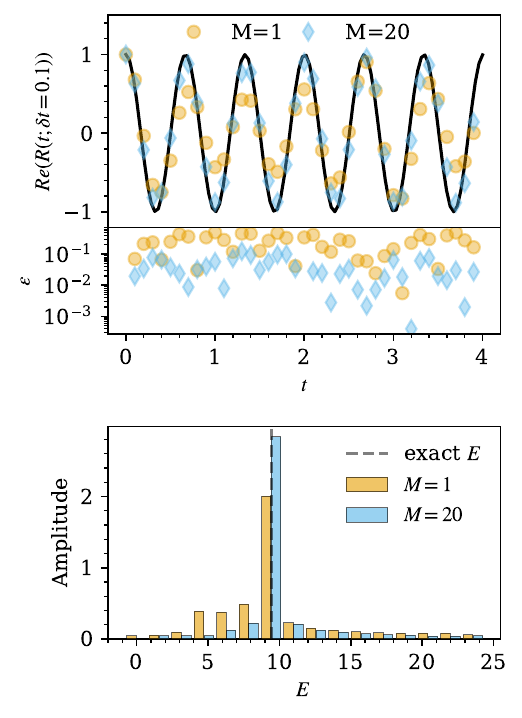}
    \caption{Analysis of real part of the recurrence probability $R(t)$ of the ground state for of $N_s=16$, $g=0.1$, $m_0=1.0$ for different $M$. Top: $R(t)$ and difference between the simulated and exact value.
    Bottom: Fourier transformation of $R(t)$.}
    \label{fig:sim}
\end{figure}

\textit{Numerical study.}
To avoid the complications involved with coupling to gauge fields, we investigate the method first utilizing the Thirring model~\cite{Thirring:1958in,Saalmann:2018tvu,Lamm:2019uyc,Banuls:2019hzc,Bakalov:2026tlk} which models fermions interacting by a four-fermion vector current-current interaction. While there are other pure fermionic models such Hubbard model, we expect the methods developed here will port over equally well and leave these other cases as a future avenue of study. The lattice Thirring model is governed by the Hamiltonian,
\begin{align}
H =&  -\frac{1}{2} \sum_{x=0}^{N-2}(\hat{a}^\dagger_x \hat{a}_{x+1} + \hat{a}^\dagger_{x+1}\hat{a}_x) + m_0 \sum_{x=0}^{N-1} (-1)^x \hat{a}^\dagger_x \hat{a}_x \notag\\
& + 2g \sum_{x=0}^{N / 2} \hat{a}^\dagger_{2x}\hat{a}_{2x} \hat{a}^\dagger_{2x+1}\hat{a}_{2x+1},
\end{align}
where we have set the lattice spacing, $a=1$, and $g$ is the coupling and $m_0$ is the bare fermion mass. 
We perform the DMRG calculations using the ITensors Library~\cite{Fishman:2020gel}. 

\begin{figure}
    \includegraphics[width=\linewidth]{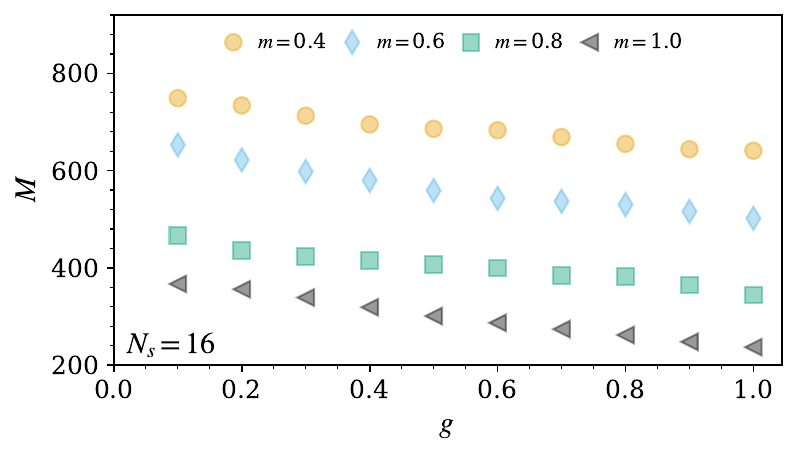}
    \caption{Values of $M$ needed to ensure $\epsilon<10^{-2}$ for $N_s=16$.}
    
    \label{fig:scale}
\end{figure}

We use as a metric of comparison the recurrence probability, also known as the Loschmidt echo,
\begin{align}
R(t) = \langle \Psi | e^{-itH} |\Psi \rangle.
\end{align}
This observable should have frequencies corresponds to eigenvalues of $H$. For this fiducial simulation we measured the 500 lowest-lying computational basis states, a Trotter step of $\delta t=0.1$, and used a Hadamard test with 10k samples to measure $R(t)$.  We show the resurgence probability in Fig. \ref{fig:sim}. We observed the anticipated improvement in accuracy with $M$ in both comparisons to exact curves and signal strength in the Fourier transformation. 

Another qualitative metric is the dependency on the number of states needed for a target integrated error,
\begin{align}
    \epsilon_{N_t} = \frac{1}{N_t} \sum_{n=1}^{N_t}|\Re(\langle \Psi| U^n(\delta t) | \Psi\rangle - \langle \phi |U^n(\delta t)|\phi\rangle)|.
\end{align}
We studied the dependence of $\epsilon(N_t)$ on $g$ and $m_0$ for $N_s\leq16$, showing the $N_s=16$ results in Fig. \ref{fig:scale}. There is a clear increase in scaling as $g,m\rightarrow0$ which corresponds to the systems with longer correlation lengths. Empirically, we find $M\propto\frac{1}{mg} \log(1/g)
    \log(1/m)$. We expect the scaling of the number of required states for a given integrated error to be approximately monotonic up to some fluctuations as the correlation length increases and strictly increasing as a function of system size, $N_s$.

\begin{figure}[!t]
    \centering
    \includegraphics[width=\linewidth]{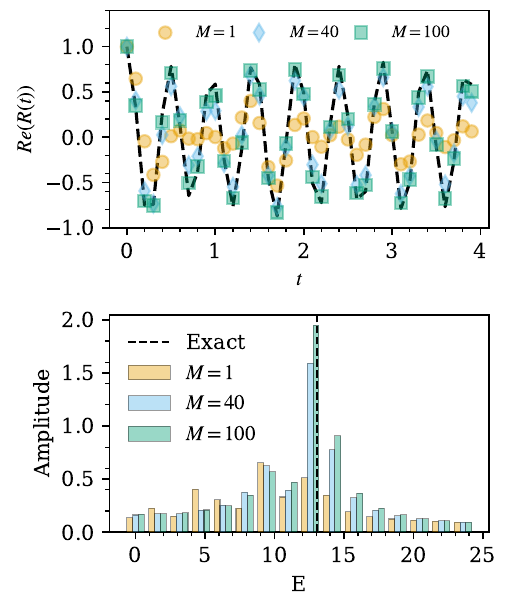}
    \caption{Analysis of first excited state with (top) $R(t)$ and (bottom) Fourier spectrum (Bottom) for the first excited state of $N_s=16$, $g=0.1$, $m_0=1.0$ for different $M$.}
    \label{fig:energetic}
\end{figure}

\begin{figure*}
\includegraphics[width=\linewidth]{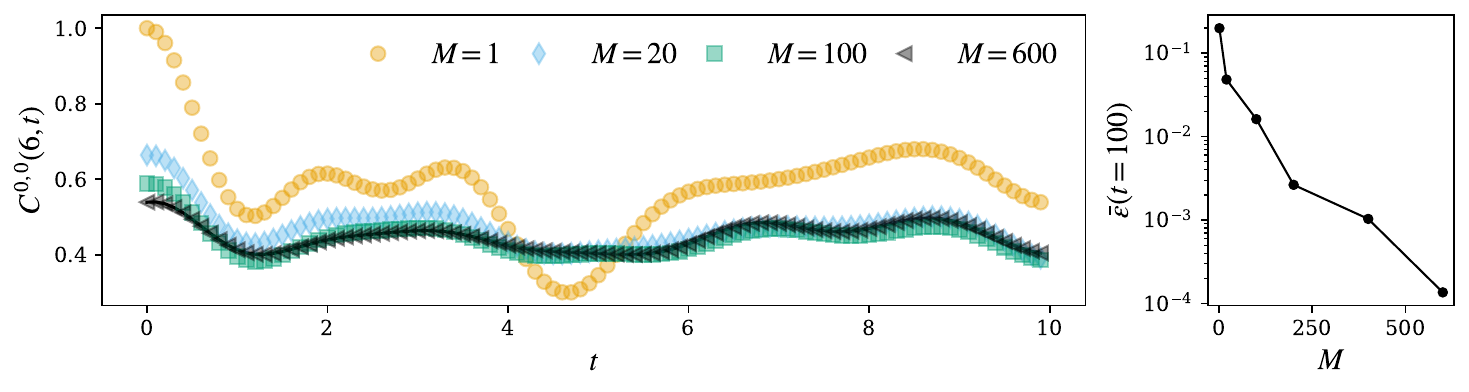}
\caption{Two-point correlation function $C^{0,0}(6, t)$ vs. $t$ and $\bar{\varepsilon}$ vs. $M$ for $m_0=0.6$, $g=0.4$, and $N=16$. Couplings are different from other examples to show case broader applicability and in different regime. }
\label{fig:currenterror}
\end{figure*}

We also investigated as proof of principle preparing the first excited state in the Thirring model, for $g=0.1$, $m_0=1.0$, $N_s=16$ as shown in Fig. \ref{fig:energetic}. While the ground state needs $M\geq 20$ states to be sufficiently accurate, the first excited state required $M\geq 100$. 

We further look at two-point correlation functions, taking as $\ket{\Psi}$ as the excited state:
\begin{align}
    C^{\mu,\nu}(x, t) = \langle \Psi | e^{itH} J^\mu(x) e^{-itH}J^\nu(0)|\Psi\rangle 
\end{align}
where the vector currents are~\cite{Lamm:2019uyc}:
\begin{align}
    J^0(x) = &\hat{a}^\dagger_x \hat{a}_x \\
    J^1(x) = & \frac{i(-1)^x}{4}\big(\hat{a}^\dagger_x(\hat{a}_{x+1} + \hat{a}_{x-1}) - h.c.\big).
\end{align}
These operators can be implemented either through numerical gradients \cite{Lamm:2019bik} or parameter-shift type rules \cite{wierichs2022general,azad2022quantum} which can be implemented with minimal circuit overhead or also utilizing LCU~\cite{Loaiza:2022btc}. Also in Fig. \ref{fig:currenterror}, we show the average cumulative error at time $t=100$.

We observed that $\bar{\varepsilon}(N,t)$ approaches a constant for late times $t>10$. Thus, with sufficient fault-tolerant resources on may load excited states onto a quantum computer and compute observables at fixed systematic uncertainty independent of the $t$ required.

\textit{Outlook.} A $\mathcal{O}(M^2)$ state preparation algorithm has been developed for fermionic theories. This method utilizes the LCU to load a weighted selection of bitstrings states computed classically onto a quantum computer. The quantum method is agnostic to the method of sampling. The required number of states depends upon physical parameters and has a polynomial scaling in volume. 

Our method is well-suited to the fault-tolerant regime for two reasons. First, it pairs naturally with phase estimation, which can extract eigenvalues with high probability, a key subroutine in computing $n$-point correlation functions. Second, as the number of retained bitstrings increases and the prepared state better approximates the true eigenstate, the overlap approaches unity, directly improving the success probability of phase estimation.

Future work in the field of LCU, approximate quantum loaders~\cite{Zhang:2026geu}, or QRAM will be fruitful. 
We expect the developments to apply to simulations in high energy physics~\cite{Lamm:2026vqz} such as studying partonic physics \cite{Li:2024nod,Chen:2025zeh,Lamm:2019uyc,Bepari:2020xqi} or scattering processes \cite{Farrell:2025nkx,Qian:2025fnx,Chai:2025qhf}. Outside of the fields of lattice gauge theories, they may be valuable for studies of condensed matter systems like Hubbard models.

\textit{Acknowledgements}
    We would like to thank Lucas Braydwood for their helpful thoughts and contributions. We acknowledge the support of the U.S. Department of Energy, Office of Science, Office of High Energy Physics Quantum Information Science Enabled Discovery (QuantISED) program ``Toward Lattice QCD on Quantum Computers" with E.G. under award number DE-SC0025940. This work was produced by Fermi Forward Discovery Group, LLC under Contract No. 89243024CSC000002 with the U.S. Department of Energy, Office of Science, Office of High Energy Physics.

\bibliography{wise}

\appendix

\section{Connection of Sampling from MPS states to \epoq}

At finite temperature a time dependent observable can be written as
\begin{align}
    \langle O(t)\rangle = \frac{\text{Tr}(e^{-\beta \hat{H}} \hat{O}(t))}{\text{Tr}(e^{-\beta \hat{H}})}.
\end{align}
In the limit that the inverse temperature $\beta\rightarrow \infty$ we find 
\begin{align}
    \lim_{\beta\rightarrow\infty} \frac{e^{-\beta \hat{H}}}{\text{Tr}(e^{-\beta \hat{H}})} = |\Omega\rangle \langle \Omega| = \sum_{i,j} \alpha_i \alpha_j^* |\psi_i\rangle \langle \psi_j |.
\end{align}
So we effectively if we can sample from a pure state either via MCMC at low temperature in the temporal direction, or directly via Hamiltonian techniques such as matrix product states, we now have a direct method to prepare "pure" states. An added benefit of this method is we can also naively see that by apply a projector that removes the ground state contribution to $\rho$, $\Pi_0 = (\mathbb{1} - |gs\rangle \langle gs|)$,
then
\begin{align}
    \lim_{\beta\rightarrow \infty } \frac{\Pi_0 e^{-\beta H} \Pi_0}{\text{Tr}(\Pi_0 e^{-\beta H} \Pi_0)} = |f.s.\rangle \langle f.s|
\end{align}
gives us the first excited state which can be sampled from as well.
In practice, however, the projector $\Pi_0$ requires knowledge of the exact ground 
state $|gs\rangle$, which is generally unavailable. Instead, we work with an approximate 
projector $\tilde{\Pi}_0 = (\mathbb{1} - |\tilde{gs}\rangle \langle \tilde{gs}|)$, where 
$|\tilde{gs}\rangle = |gs\rangle + |\delta\rangle$ and $|\delta\rangle$ is the error in 
the ground state approximation. This imperfect projection means the resulting state retains 
a residual ground state admixture. The error introduced by this approximation is of the same 
character as the systematic error analyzed in the following section: it is bounded by the 
overlap $|\langle \delta | gs \rangle|$ and is directly improvable by increasing the 
bond dimension or number of retained states in the classical approximation. We therefore 
treat the quality of the approximate projector as part of the same systematic error 
framework developed below, rather than a fundamental obstruction.

\end{document}